\documentclass{article}

\PassOptionsToPackage{numbers, compress, square}{natbib}

\usepackage[final]{neurips_2019}

\usepackage[utf8]{inputenc} 
\usepackage[T1]{fontenc}    
\usepackage{hyperref}       
\usepackage{url}            
\usepackage{booktabs}       
\usepackage{amsfonts}       
\usepackage{nicefrac}       
\usepackage{microtype}      

\usepackage{mathptmx}

\usepackage{graphicx}

\usepackage{subcaption}
\title{Hate Speech in Pixels: Detection of Offensive Memes towards Automatic Moderation}

\author{
Benet Oriol Sabat, Cristian Canton Ferrer, Xavier Giro-i-Nieto\\
Universitat Politecnica de Catalunya - UPC\\
}
\begin{document}

\maketitle

\begin{abstract}
This work addresses the challenge of hate speech detection in Internet memes, and attempts using visual information to automatically detect hate speech, unlike any previous work of our knowledge. Memes are pixel-based multimedia documents that contain photos or illustrations together with phrases which, when combined, usually adopt a funny meaning. 
However, \textit{hate memes} are also used to spread hate through social networks, so their automatic detection would help reduce their harmful societal impact.
Our results indicate that the model can learn to detect some of the memes, but that the task is far from being solved with this simple architecture.
While previous work focuses on linguistic hate speech, our experiments indicate how the visual modality can be much more informative for hate speech detection than the linguistic one in memes.
In our experiments, we built a dataset of 5,020 memes to train and evaluate a multi-layer perceptron over the visual and language representations, whether independently or fused.


\end{abstract}

\section{Motivation}

\par The spread of misinformation or hate messages through social media is a central societal challenge given the unprecedented broadcast potential of these tools. While there already exist some moderation mechanisms such as crowd-sourced abuse reports and dedicated human teams of moderators, the huge and growing scale of these networks requires some degree of automation for the task.

\par Social networks have already introduced many tools to detect offensive or misleading content, both for visual and textual content, ranging from nudity and pornography \cite{Karamizadeh2018, 8614362} to hate speech \cite{Fortuna2018} and misinformation \cite{Bouchard2019}. However, machine learning is still facing some challenges when processing borderline or figurative content such as nudity in paintings, political satire or other forms of humorous content. In particular for the case of hate speech, rapidly evolving topics and shifting trends in social media make its detection a topic of constant and active research.

\par This work takes one step forward and instead of focusing on visual or linguistic content alone, we tackle the challenging problem of detecting hate speech in memes. Memes are a form of humorist multimedia document which is normally based on an image with some sort of caption text embedded in the image pixels. Memes have gained a lot of popularity in the last few years and have been used in many different contexts, specially by young people. However, this format has also been used to produce and disseminate hate speech in the form of dark humour. The multimodal nature of memes makes it very challenging to analyze because, while the visual and linguistic information is typically neutral or actually funny in isolation, their combination may result in hate speech messages.

\par Our work explores the potential of state of the art deep neural networks to detect hate speech in memes. We study the gain in accuracy when detecting hate speech in memes by fusing the vision and language representations, when compared with the two modalities apart. Our  experiments indicate that while meme detection is a multimodal problem that benefits by analyzing both modalities, this societal task is far from being solve given the high abstraction level of the messages contained in memes.

\section{Related Work}
\par Hate speech is a widely studied topic in the context of social science. This phenomena has been monitored, tracked, measured or quantified in a number of occasions \cite{hate1,hate2,hate3}. It appears in media such as newspapers or TV news, but one of the main focus of hate speech with very diverse targets has appeared in social networks \cite{hate_mesure,hate_twitter1,hate_twitter2}.
Most works in hate speech detection has focused in language. The most common approach is to generate an embedding of some kind, using bag of words \cite{hate_twitter1} or N-gram features \cite{hate_text1} and many times using expert knowledge for keywords. After that, the embedding is fed to a binary classifier to predict hate speech. Up to our knowledge, there is no previous work on detecting hate speech when combining language with visual content as in memes. Our technical solution is inspired by \cite{blandfort2019multimodal} in which gang violence on social media was predicted from a multimodal approach that fused images and text. Their model extracted features from both modalities using pretrained embeddings for language and vision, and later merged both vectors to feed the multimodal features into a classifier.
\section{Model}
\par
\begin{figure}
  \centering
  \includegraphics[width=\textwidth]{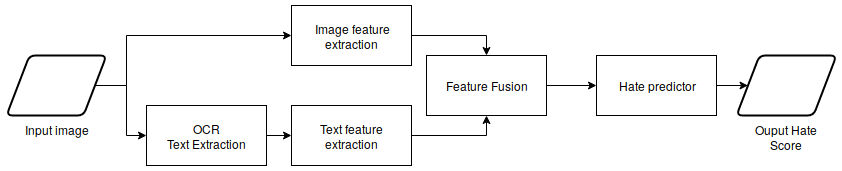}
  \caption{Block diagram of the system}
  \label{fig:overall}
\end{figure}

 The overall system expects an Internet meme input, and produces a hate score as an output.  Figure \ref{fig:overall} shows a block diagram of the proposed solution.
The first step of the process is extracting the text of the image with Optical Character Recognition (OCR).  The text detected by the OCR is encoded in a BERT \cite{bert} representation for language. 
We used the Tesseract 4.0.0 OCR \footnote{\url{https://github.com/tesseract-ocr/tesseract}} with a Python wrapper\footnote{\url{https://pypi.org/project/pytesseract/}} on top.
This encoding generates contextual (sub)words embeddings, which we turn into a sentence embedding by averaging them.
We used a PyTorch implementation available at the repo below \footnote{\url{https://github.com/huggingface/pytorch-pretrained-BERT}}. 
This implementations includes multiple pretrained versions and we chose the one called \textit{bert-base-multilingual-cased}. 
This version has 12 layers, 768 hidden dimensions, 12 attention heads with a total of 110M parameters and is trained on 104 languages.

The visual information was encoded with a VGG-16 convolutional neural network \cite{simonyan2014very}, trained on ImageNet \cite{deng2009imagenet}.
Then we used the activations from a hidden layer as feature vectors for the image,
Specifically, we used the last hidden before output, which has 4096 dimensions.
We obtained the pretrained model from the TorchVision module in PyTorch.

The text and image encodings were combined by concatenation, which resulted in a feature vector of 4,864 dimensions. 
This multimodal representation was afterward fed as input into a multi-layer perceptron (MLP) with two hidden layer of 100 neurons with a ReLU activation function.
The last single neuron with no activation function was added at the end to predict the hate speech detection score.

\section{Dataset}
\label{sec:datasets}

We built a dataset for the task of hate speech detection in memes with 5,020 images that were weakly labeled into \textit{hate} or \textit{non-hate} memes, depending on their source. \textit{Hate} memes were retrieved from Google Images with a downloading tool\footnote{\url{https://github.com/hardikvasa/google-images-download}}. We used the following queries to collect a total of 1,695 \textit{hate memes}: \textit{racist meme} (643 memes), \textit{jew meme} (551 memes), and \textit{muslim meme} (501 Memes). \textit{Non-hate} memes were obtained from the Reddit Memes Dataset \footnote{\url{https://www.kaggle.com/sayangoswami/reddit-memes-dataset}}. We assumed that all memes in the dataset do not contain any hate message, as we considered that average Reddit memes do not belong to this class. A total of 3,325 non-hate memes were collected.  We split the dataset into train (4266 memes) and validation (754 memes) subsets. The splits were random and the distribution of classes in the two  subsets is the same. We didn't split the dataset into three subsets because of the small amount of data we had and decided to rely on the validation set metrics.
    



\section{Experiments}

Our experiments aimed at estimating the potential of a multimodal hate speech detector, and study the impact of a multimodal analysis when compared to using language or vision only. 

We estimated the parameters of the MLP on top of the encoding of the meme with an ADAM optimizer with a lr=0.1, betas=(0.9, 0.999) and $\varepsilon=10^{-8}$, weight decay=0, a batch size of 25, and a drop out of 0.2 on the first hidden layer.
The network was trained with a a Mean Squared Error (MSE) loss, but assessed in terms of binary accuracy. 



\begin{figure}[h!]
  \centering
  \begin{subfigure}[b]{0.32\linewidth}
    \includegraphics[width=\linewidth]{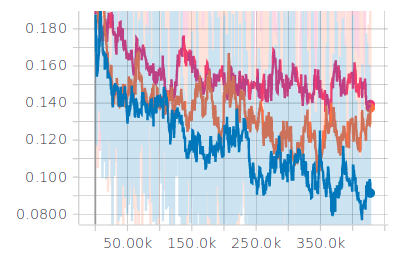}
    \caption{}
  \end{subfigure}
  \begin{subfigure}[b]{0.32\linewidth}
    \includegraphics[width=\linewidth]{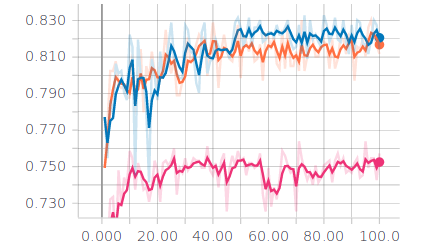}
    \caption{}
  \end{subfigure}
   \begin{subfigure}[b]{0.32\linewidth}
    \includegraphics[width=\linewidth]{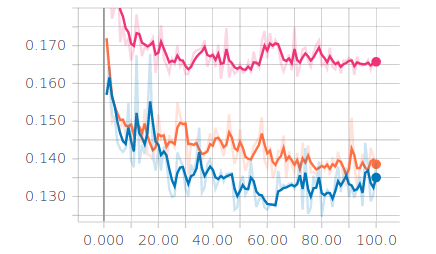}
    \caption{}
  \end{subfigure}
  \caption{Results refer to the \textit{language only} (red), \textit{vision only} (orange) and \textit{multimodal} (blue) solutions. In (a) we see the train loss, in (b) we see the validation loss and in (c) the validation average binary accuracy. The x-axis is \#batch for (a) and \#epoch for (b) and (c).}
  \label{fig:multimodal_curves}
\end{figure}

Figure \ref{fig:multimodal_curves} presents the results of the training with the three considered configurations: language only, vision only, and a multimodal solution.
In the single modality cases, the input layer of the MLP is adjusted to the size of the encoded representation.
The curves show how the blue line representing the \textit{multimodal} case obtains the best results, closely followed by the orange one of the \textit{vision only} case.
The \textit{language only} configuration performs clearly worse than the other two.
Nevertheless, the three curves are consistenly over the baseline accuracy of $0.66$, which would be achieved by a dummy predictor of \textit{Non-hate} class, because of the 34\%-66\% class imbalance of the dataset.


\begin{table}
  \caption{Accuracy results for the three configurations}
  \label{results-table}
  \centering
  \begin{tabular}{ccc}
    \toprule
    \cmidrule(r){1-2}
    Model     & Max. Accuracy  & Smth. Max. Accuracy\\
    \midrule
    Multimodal & 0.833  & 0.823     \\
    Image     & 0.830 & 0.804      \\
    Text     & 0.761 & 0.750  \\
    \bottomrule
  \end{tabular}
  \label{tab:accuracies}
\end{table}

Table \ref{tab:accuracies} provides numerical results comparing the three configurations based on two different metrics: \textit{Max. Accuracy} corresponds to the best accuracy obtained in any epoch, while \textit{Smth Max. Accuracy} corresponds to the smoothed accuracy to which the model was converging. This was estimated by smoothing the curve with a momentum average and picking the best value. We thought the second metric was a good estimation of the real performance of the model due to the huge validation accuracy fluctuation between epochs in evaluation. Also, since the classes are imbalanced, we computed the precision-recall curve for the best multimodal model, getting an Average Precision of $0.81$.



\begin{figure}[h!]
  \centering
  \begin{subfigure}[t]{0.25\linewidth}
    \includegraphics[width=\linewidth]{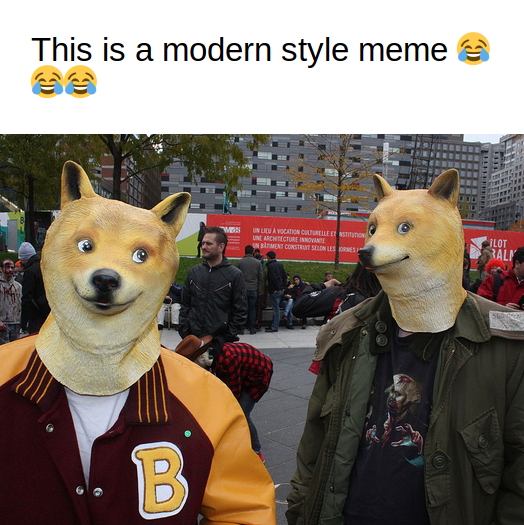}
    \caption{Modern meme}
  \end{subfigure}
  \begin{subfigure}[t]{0.25\linewidth}
    \includegraphics[width=\linewidth]{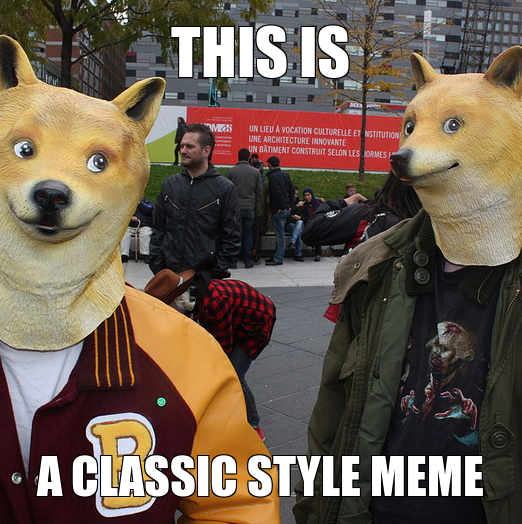}
    \caption{Classic Meme}
  \end{subfigure}
  \begin{subfigure}[t]{0.25\linewidth}
    \includegraphics[width=\linewidth]{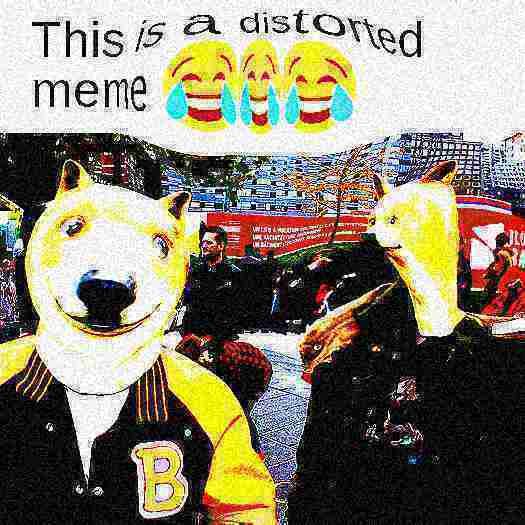}
    \caption{Distorted Meme}
  \end{subfigure}
  \caption{Three examples of the diffenent meme styles}
  \label{fig:meme_examples}
\end{figure}

We consider that the superior performance of the \textit{vision only} configuration over the \textit{language only} one may be due to a diversity of reasons.
 Firstly, the most obvious one is that the dimensionality of the image representation (4096) is much larger than the linguistic one (768), so it has the capacity to encode more information. Also, the different models have different number of parameters due to different MLP input and we didn't take into consideration this variation of the model's capacity. Secondly, we think there might a visual bias on the dataset. Mainly, because there are more \textit{modern style memes} on the no hate class and more \textit{classic style memes} in the hate class. \textit{Classic} or \textit{modern} memes refer basically to the format and placement of the text. Figure \ref{fig:meme_examples} (a) and (b) are examples of them. 
 Also, we found some false positives in the hate class and there might be false negatives in the non-hate Reddit set.
Finally, memes are often highly compressed images with an important level of distortion. This fact may affect the quality of the OCR recognition and, therefore, the language encoding, as shown in the Figure \ref{fig:meme_examples} (c). 

The training code and models are publicly available to facilitate reproducibility. \footnote{\url{https://github.com/imatge-upc/hate-speech-detection}}

\section{Conclusions}

Our study on hate speech detection in memes concludes that it is possible to automatize the task, in the sense that a simple configuration using state of the art image and text encoders can detect some of them. 
However, the problem is far from being solved, because the best accuracy obtained of $0.83$ seems modest despite being much better than greedy solution of predicting always the most frequent class.
The proposed system may be used for filtering some of the memes distributed through a social network, but it would still require a human moderator for many of them.

Unfortunately, the system may actually also be used for the opposite of detecting hate speech memes, but helping in their creation.
Given a large amount of sentences and images, a misuse of the system may assess the hate score of each possible pair of text and image to find novel combinations with an expect high hate level.

The experiments also show that the visual cues are much more important than the linguistic ones when detecting hate speech memes, a totally opposite scenario to previous studies focusing on language-based hate speech detection.
While the best results are obtained with the multimodal approach, the gain with respect to the vision only one is small.
A practical deployment of this system should evaluate whether the computation cost of running the OCR and encoding the extracted text is worthy based on the reported gains in accuracy.

The present work poses a new challenge to the multimedia analysis community, which has been proven to be difficult but not impossible. 
Given the rich affective and societal content in memes, an effective solution should probably also take into account much more additional information than just the one contained in the meme, such as the societal context in which the meme is posted.


\newpage 

\section*{Acknowledgements}
This work has been developed in the framework of project TEC2016-75976-R, funded by the Spanish Ministerio de Economía y Competitividad and the European Regional Development Fund (ERDF), and the Industrial Doctorate 2017-DI-011 funded by the Government of Catalonia. We gratefully acknowledge the support of NVIDIA Corporation with the donation of some of the GPUs used for this work.

\bibliography{refs} 
\bibliographystyle{abbrvnat}
\end{document}